\documentclass[twocolumn,aps,prl,showkeys,amsmath,amssymb]{revtex4}

\usepackage{graphicx}
\usepackage{dcolumn}
\usepackage{amsmath}%
\usepackage{amssymb}%

\begin{document}



\title{Derivation of the Maxwell's Equations Based on
a Continuum Mechanical Model of Vacuum and a Singularity Model of
Electric Charges}

\author{Xiao-Song Wang}
\email{wangxs1999@yahoo.com}
\affiliation{State Key Laboratory of Nonlinear Mechanics (LNM),
Institute of Mechanics, Chinese Academy of Sciences, Beijing,
100080, China}
\date{\today}

\begin{abstract}
    We speculate that the universe may be filled with a visco-elastic
continuum which may be called aether. Thus, the Maxwell's equations
in vacuum are derived by methods of continuum mechanics based on a
continuum mechanical model of vacuum and a singularity model of
electric charges.
\end{abstract}


\maketitle

\maketitle \setcounter{section}{0} \stepcounter{section}
\section{\thesection . \ \label{Introduction}Introduction}
\thispagestyle{plain}

\newtheorem{hypothesis}{Hypothesis}
\newtheorem{definition}[hypothesis]{Definition}

The Maxwell's equations in vacuum can be written as
\cite{Jackson1963}

\begin{eqnarray}
\nabla \cdot \mbox{\upshape\bfseries{E}}&=&\frac{\rho_{e}}{\epsilon_{0}}, \label{The Maxwell's equations 1-10} \\
 \nabla \times\mbox{\upshape\bfseries{E}} &=&
 -\frac{\partial \mbox{\upshape\bfseries{B}}}{\partial t}, \label{The Maxwell's equations 1-20} \\
 \nabla \cdot \mbox{\upshape\bfseries{B}}&=& 0, \label{The Maxwell's equations 1-30} \\
 \frac{1}{\mu_{0}}\nabla \times\mbox{\upshape\bfseries{B}}
 &=& \mbox{\upshape\bfseries{j}}+\epsilon_{0}\frac{\partial \mbox{\upshape\bfseries{E}}}{\partial t} , \label{The Maxwell's equations 1-40}
\end{eqnarray}
where $\mbox{\upshape\bfseries{E}}$ is the electric field vector,
$\mbox{\upshape\bfseries{B}}$ is the magnetic induction vector,
$\rho_{e}$ is the density field of electric charges, $\epsilon_{0}$
is the dielectric constant of vacuum, $\mu_{0}$ is magnetic
permeability of vacuum, $t$ is time, $\nabla =
\mbox{\upshape\bfseries{i}}\frac{\partial }{\partial x} +
\mbox{\upshape\bfseries{j}}\frac{\partial }{\partial y} +
\mbox{\upshape\bfseries{k}}\frac{\partial }{\partial z}$ is the
Hamilton operator.

  The main purpose of this paper is to derive the above mentioned
  Maxwell's equations in vacuum based on
a continuum mechanics model of vacuum and a singularity model of
electric charges.

  The motive of this paper is to seek a mechanism of
electromagnetic phenomena. The reasons why new mechanical models
 of electromagnetic field are interesting may be summarized as follows.

    Firstly, there exists some electromagnetic phenomena that could not
be interpreted by the present theories of electromagnetic field,
e.g., the spin of electrons \cite{Jackson1963,Landau-Lifshitz1958},
the Aharonov-Bohm effect\cite{Aharonov-Bohm1959,Chambers1960}. New
theories of of electromagnetic phenomena may view these problems
from new angles.

   Secondly, there exists some inconsistencies and inner difficulties in the  classical
   electrodynamics, e.g., the inadequacy of Li\'{e}enard-Wiechert potentials
   \cite{Landau-Lifshitz1975,Whitney1988,Chubykalo-Smirnov-Rueda1996}. New theories of electromagnetic phenomena
    may overcome such difficulties.

    Thirdly, there exists some divergence problems in quantum
   electrodynamics\cite{Dirac1978}. In Dirac's words,
   'I must say that I am very dissatisfied with the situation,
   because this so-called good theory does involve neglecting infinities
   which appear in its equations, neglecting them in an arbitrary way.
   This is just not sensible mathematics.' New theories of electromagnetic phenomena
    may open new ways to solve such problems.

   Fourthly, since quantum theory shows that vacuum is not empty and
has physical effects, e.g., the Casimir
effect\cite{Lamoreaux2005,Intravaia-Lambrecht2005,Guo-Zhao2004,Davies2005},
 it is valuable to reexamine the old concept of electromagnetic aether.

   Fifthly, from the point view of reductionism, Maxwell's theory of electromagnetic field
  can only be regarded as a phenomenological theory. Although
Maxwell's theory is a field theory, the concept of field is
different from that of continuum mechanics
\cite{Truesdell1966,Fung1977,Eringen1980,Landau-Lifshitz1987}
because of the absence of a continuum. Thus, from the point view of
reductionism, the mechanism of electromagnetic phenomena remains an
unsolved problem in physics \cite{Whittaker1953}.

   Sixthly, one of the puzzles in physics is the problem
 of dark matter and dark energy (refer to, for instance, \cite{Ellis2003,Linder2004,Bernardeau2003,Bergstrom2005,Beacom2005,Akerib2006,Feng2006,Fayet2006,Carena2006}).
 New theories of electromagnetic phenomena may provide new ideals to attack this
problem.

   Finally, one of the tasks of physics is the unification of the four
fundamental interactions in the universe. New theories of
electromagnetic phenomena may shed some light on this puzzle.

   To conclude, it seems that new considerations on electromagnetic
phenomena is needed. It is worthy keeping an open mind with respect
to all the theories of electromagnetic phenomena before the above
problems been solved.

   Now let us briefly review the long history of the mechanical
interpretations of electromagnetic phenomena.

   According to E. T.
Whittaker\cite{Whittaker1953}, Descartes was the first to bring the
concept of the aether into science by suggesting that it has
mechanical properties according to E. T.
Whittaker\cite{Whittaker1953}. Descartes believed that every
physical phenomenon could be interpreted in the construction of a
mechanical model of the universe. William Watson and Benjamin
Franklin introduced the one-fluid theory of electricity
independently in 1746 \cite{Whittaker1953}. Henry Cavendish
attempted to explain some of the principal phenomena of electricity
by means of an elastic fluid in 1771\cite{Whittaker1953}. Not
contented with the above mentioned one-fluid theory of electricity,
du Fay, Robert Symmer and C. A. Coulomb developed a two-fluid theory
of electricity from 1733 to 1789 \cite{Whittaker1953}.

Before the unification of electromagnetic phenomena and light
phenomena by Maxwell in 1860s, light phenomena were also studied
independently based on Descartes' scientific research program of the
mechanical theory of nature. John Bernoulli introduced a fluidic
aether theory of light in 1752\cite{Whittaker1953}. Euler believed
that all electrical phenomena is caused by the same aether that
propagates light. Furthermore, Euler attempted to
 explain gravity in terms of his single fluidic aether\cite{Whittaker1953}.

In 1821, in order to explain polarisation of light, A. J. Frensnel
proposed an aether model which is able to transmit transverse waves.
After the advent of this successful transverse wave theory of light
of Frensnel, those imponderable fluid theories were abandoned.
Frensnel's dynamical theory of a luminiferous aether had an
important influence on the mechanical theories of nature in the
nineteenth century \cite{Whittaker1953}. Inspired by Frensnel's
luminiferous aether theory, numerous dynamical theories of elastic
solid aether were established by Stokes, Cauchy, Green, MacCullagh,
Boussinesq, Riemann and William Thomson, see, for instance,
\cite{Whittaker1953}.

Thomson's analogies between electrical phenomena and elasticity
helped to inspire James Clark Maxwell established a mechanical model
of electrical phenomena \cite{Whittaker1953}. Strongly impressed by
Faraday's theory of lines of forces, Maxwell compared the Faraday's
lines of forces with the lines of flow of a fluid. In 1861, in order
to obtain a mechanical interpretation of electromagnetic phenomena,
Maxwell established a mechanical model of a magneto-electric medium.
Maxwell's magneto-electric medium is a cellular aether, looks like a
honeycomb. Each cell of the aether consists of a molecular vortex
surrounded by a layer of idle-wheel particles. In a remarkable paper
published in 1864, Maxwell established a group of equations, which
were named after his name later, to describe the electromagnetic
phenomena.

In 1878, G. F. FitzGerald compared the magnetic force and the
velocity in a quasi-elastic solid of the type first suggested by
MacCullagh \cite{Whittaker1953}. FitzGerald's mechanical model of
the electromagnetic aether were studied by A. Sommerfeld, by R.
Reiff and by Sir J. Larmor later \cite{Whittaker1953}.

Because of some dissatisfactions with the mechanical models of the
electromagnetic aether and the success of the theory of
electromagnetic field, the mechanical world view was replaced by the
electromagnetic world view gradually. Therefore, the concepts of a
luminiferous aether and an elastic solid aether were replaced by the
concepts of a electromagnetic aether or the electromagnetic field.
This paradigm shift in scientific research was attributed to many
scientists, including Faraday, Maxwell, Sir J. Larmor, H. A.
Lorentz, J. J. Thomson, H. R. Hertz, Ludwig Lorenz, Emil Wiechert,
Paul Drude, Wilhelm Wien, etc., see, for instance,
\cite{Whittaker1953}.

    In a remarkable paper published in 1905,
Einstein abandoned the concept of aether\cite{Einstein1905}.
However, Einstein's assertion did not cease the exploration of
aether (refer to, for instance,
\cite{Whittaker1953,Golebiewska-Lasta1979,Barut1988,Oldershaw1989a,Oldershaw1989b,Dmitriyev1992,Larson1998,Marmanis1998,Dmitriyev1998,Zareski2001,Dmitriyev2003}).
Einstein changed his attitude later and introduced his new concept
of ether\cite{Einstein1920,Kostro2000}. In 1979, A. A.
Golebiewska-Lasta observed the similarity between the
electromagnetic field and the linear dislocation field
\cite{Golebiewska-Lasta1979}. V. P. Dmitriyev have studied the
similarity between the electromagnetism and the linear elasticity
since
1992\cite{Dmitriyev1992,Dmitriyev1998,Dmitriyev2003,Dmitriyev2004}.
H. Marmanis established a new theory of turbulence based on the
analogy between electromagnetism and turbulent hydrodynamics in
1998\cite{Marmanis1998}. D. J. Larson derived the Maxwell's
equations from a simple two-component solid-mechanical aether in
1998 \cite{Larson1998}. D. Zareski gave an elastic interpretation of
electrodynamics in 2001 \cite{Zareski2001}. I regret to admit that
it is impossible for me to mention all the
 works related to this field in history.

  Inspired by the above mentioned works, we
show that the Maxwell's equations of electromagnetic field can be
derived based on a continuum mechanics model of vacuum and a
singularity model of electric charges.

\stepcounter{section}
\section{\thesection . \  \label{dimensional analysis}Clues obtained from dimensional analysis}

According to Descartes' scientific research program which was based
on his mechanical view of nature, the electromagnetic phenomenon
must be and can be interpreted based on the mechanical motions of
aether particles.

Therefore, all the physical quantities appearing in the theory of
electromagnetic field must be mechanical quantities.

Thus, in order to establish a successful mechanical model of
electromagnetic field, we should undertake a careful dimensional
analysis (refer to, for instance, \cite{Tan2005}), of electric field
vector $\mbox{\upshape\bfseries{E}}$, magnetic induction vector
$\mbox{\upshape\bfseries{B}}$, the density field of electric charges
$\rho_{q}$, the dielectric constant of vacuum $\epsilon_{0}$,
magnetic permeability of vacuum $\mu_{0}$, etc.

It is known that the Maxwell's equations (\ref{The Maxwell's
equations 1-10}-\ref{The Maxwell's equations 1-40}) in vacuum can
also be expressed as \cite{Jackson1963}
\begin{eqnarray}
&&\nabla^2 \phi + \frac{\partial
}{\partial t} (\nabla \cdot \mbox{\upshape\bfseries{A}}) = - \frac{\rho_{e}}{\epsilon_{0}},\label{Maxwell's equations 2-1}\\
&&\nabla^2 \mbox{\upshape\bfseries{A}} - \nabla(\nabla \cdot
 \mbox{\upshape\bfseries{A}}) - \mu_{0}\epsilon_{0}\frac{\partial
}{\partial t}\left ( \nabla \phi - \frac{\partial
\mbox{\upshape\bfseries{A}}}{\partial t}\right )= -
\mbox{\upshape\bfseries{j}},\label{Maxwell's equations 2-2}
\end{eqnarray}
where $\phi$ is the scalar electromagnetic potential,
$\mbox{\upshape\bfseries{A}}$ is the vector electromagnetic
potential, $\nabla^2 = \frac{\partial^2 }{\partial x^2} +
\frac{\partial^2 }{\partial y^2} + \frac{\partial^2 }{\partial z^2}$
is the Laplace operator.

Noticing the similarity between the Eq.(\ref{Maxwell's equations
2-2}) and the equation (\ref{The vector form of the equation of
momentum conservation}) of momentum conservation of elastic solids,
it is natural to speculate that the dimension of the vector
electromagnetic potential $\mbox{\upshape\bfseries{A}}$ of aether
has the same of the dimension of the displacement vector
$\mbox{\upshape\bfseries{u}}$ of an elastic solid.

In 1846, W. Thomson compared electric phenomena with elasticity. He
pointed out that the elastic displacement
$\mbox{\upshape\bfseries{u}}$ of an incompressible elastic solid is
an possible analogy with the vector electromagnetic potential
$\mbox{\upshape\bfseries{A}}$ \cite{Whittaker1953}.

Inspired by this clue, let us set out to investigate in this
direction further in the following sections.

\stepcounter{section}
\section{\thesection . \  \label{Vacuum}a visco-elastic continuum model of vacuum}

The purpose of this section is to establish a visco-elastic
continuum mechanical model of vacuum.

In 1845-1862, Stokes suggested the aether might behave like a
glue-water jelly\citep{Stokes1845,Stokes1849,Stokes1862}. He
believed that the aether would act like a fluid with respect to
translational motion of large bodies through it, but would still
posses elasticity to produce small transverse vibrations.

Following Stokes, we propose a visco-elastic continuum model of
vacuum.

\begin{hypothesis}\label{aether}
Suppose the universe is filled with a kind of continuously
distributed material.
\end{hypothesis}

This material may be called aether for convenience.

In order to establish a continuum mechanical theory of aether, we
must introduce some hypotheses based on experimental data of the
macroscopic behavior of vacuum.

\begin{hypothesis}
We suppose that the all the mechanical quantities of the aether
under consideration, such as density, displacements, strains,
stresses, etc., are piecewise continuous functions of space and
time. Furthermore, we suppose that the material points of the aether
remain one-to-one correspondence with the material points before
deformation happens.
\end{hypothesis}

\begin{hypothesis}\label{homogeneous hypothesis 3-10}
We suppose that the material of the aether under consideration is
homogeneous, that is
\begin{math}
\frac{\partial\rho}{\partial x}=\frac{\partial\rho}{\partial y}
 =\frac{\partial\rho}{\partial z}=\frac{\partial\rho}{\partial t}=0,
\end{math}
where $\rho$ is the density of aether.
\end{hypothesis}

\begin{hypothesis}
Suppose that the deformation processes of aether are isothermal
processes. We neglect the thermal effects.
\end{hypothesis}

\begin{hypothesis}
Suppose the deformation processes is not influenced by the gradient
of stress tensor.
\end{hypothesis}

\begin{hypothesis}
We suppose that the material of the aether under consideration is
isotropic.
\end{hypothesis}

\begin{hypothesis}
We suppose that the deformaton of the aether under consideration is
small.
\end{hypothesis}

\begin{hypothesis}
We suppose that there are no initial stress and strain in the body
under consideration.
\end{hypothesis}

When aether is subjected to a set of external forces, the relative
positions of the aether particles forming the body changes.

In order to described the deformation of the aether, let us
introduce a Cartesian coordinate system $\{ o, x, y, z \}$ or $\{ o,
x_1, x_2, x_3 \}$ which is static relative to the aether. Now we may
introduce the definition of displacement vector
$\mbox{\upshape\bfseries{u}}$ of every point in the aether as
\begin{equation}
\mbox{\upshape\bfseries{u}} = \mbox{\upshape\bfseries{r}} -
\mbox{\upshape\bfseries{r}}_{0},
\end{equation}
where $\mbox{\upshape\bfseries{r}}_{0}$ is the position of the point
before the deformation, $\mbox{\upshape\bfseries{r}}$ is the
position of the point after the deformation.

The displacement vector may be written as
$\mbox{\upshape\bfseries{u}}
=u_1\mbox{\upshape\bfseries{i}}+u_2\mbox{\upshape\bfseries{j}}+u_3\mbox{\upshape\bfseries{k}}$
 or $\mbox{\upshape\bfseries{u}}=u\mbox{\upshape\bfseries{i}}+v\mbox{\upshape\bfseries{j}}+w\mbox{\upshape\bfseries{k}}$,
where $\mbox{\upshape\bfseries{i}}$, $\mbox{\upshape\bfseries{j}}$,
$\mbox{\upshape\bfseries{k}}$ are the three unit vectors along the
three coordinates.

The gradient of the displacement vector
$\mbox{\upshape\bfseries{u}}$ is the relative displacement tensor
$u_{i,j} =\frac{\partial u_i}{\partial x_j}$.

We can decompose the tensor  $u_{i,j}$ into two parts, symmetric
$\varepsilon_{ij}$ and skew-symmetric $\omega_{ij}$ (refer to, for
instance, \cite{Eringen1975,Fung1977,Landau-Lifshitz1986}).
\begin{equation}\label{decompose}
 u_{i,j} = \frac{1}{2}(u_{i,j}+u_{j,i}) + \frac{1}{2}(u_{i,j} - u_{j,i})
 = \varepsilon_{ij} + \omega_{ij}
\end{equation}

\begin{alignat}{2}\label{strain and skew}
\varepsilon_{ij}
 &=\frac{1}{2}(u_{i,j}+u_{j,i}),
    &\hspace{30pt}   \omega_{ij}
 &=\frac{1}{2}(u_{i,j} - u_{j,i}).
\end{alignat}

The symmetric tensor $\varepsilon_{ij}$ represents pure deformation
of the body at a point and is called  strain tensor (refer to, for
instance, \cite{Eringen1975,Fung1977,Landau-Lifshitz1986}). The
matrix form and the indicial notation of strain tensor
$\varepsilon_{ij}$ are
\begin{equation}\label{matrix form of small strain tensor}
 \varepsilon_{ij} = \begin{pmatrix}
   \frac{\partial u}{\partial x}
 & \frac{1}{2}\left (\frac{\partial u}{\partial y} + \frac{\partial v}{\partial x}\right )
 & \frac{1}{2}\left (\frac{\partial u}{\partial z} + \frac{\partial w}{\partial x}\right )\\
   \frac{1}{2}\left (\frac{\partial v}{\partial x} + \frac{\partial u}{\partial y}\right )
 & \frac{\partial v}{\partial y})
 & \frac{1}{2}\left (\frac{\partial v}{\partial z} + \frac{\partial w}{\partial y}\right )\\
   \frac{1}{2}\left (\frac{\partial w}{\partial x} + \frac{\partial u}{\partial z}\right )
 & \frac{1}{2}\left (\frac{\partial w}{\partial y} + \frac{\partial v}{\partial z}\right )
 & \frac{\partial w}{\partial z}
 \end{pmatrix}
\end{equation}

The strain-displacements equations can be obtained from
Eq.(\ref{matrix form of small strain tensor})
\begin{equation}\label{strain-displacements}
\begin{aligned}
 \varepsilon_{11}  = \frac{\partial u}{\partial x},
 &\hspace{20pt} & \varepsilon_{12} = \varepsilon_{21}
  = \frac{1}{2}\left (\frac{\partial u}{\partial y} + \frac{\partial v}{\partial x}\right ),\\
 \varepsilon_{22}  = \frac{\partial v}{\partial y},
 &\hspace{20pt} & \varepsilon_{23} = \varepsilon_{32}
  = \frac{1}{2}\left (\frac{\partial v}{\partial z} + \frac{\partial w}{\partial y}\right ),\\
 \varepsilon_{33} = \frac{\partial w}{\partial z},
 &\hspace{20pt} & \varepsilon_{31} = \varepsilon_{13}
  = \frac{1}{2}\left (\frac{\partial w}{\partial x} + \frac{\partial u}{\partial z}\right ).
\end{aligned}
\end{equation}

For convenience we can introduce the definitions of mean strain
$\varepsilon_m$ and strain deviator $e_{ij}$ as
\begin{equation}\label{mean strain}
 \varepsilon_m
 = \frac{1}{3}(\varepsilon_{11} + \varepsilon_{22} +
 \varepsilon_{33}),
\end{equation}

\begin{equation}\label{small strain deviator}
 e_{ij} =\varepsilon_{ij} - \varepsilon_m
 =
 \begin{pmatrix}
\varepsilon_{11} - \varepsilon_m
 & \varepsilon_{12}
 & \varepsilon_{13}\\
\varepsilon_{21}
 & \varepsilon_{22} - \varepsilon_m
 & \varepsilon_{23}\\
\varepsilon_{31}
 & \varepsilon_{32}
 & \varepsilon_{33} - \varepsilon_m
 \end{pmatrix}.
\end{equation}

When the aether is deformed, internal forces arise due to the
deformation. The the indicial notation of the stress tensor
$\sigma_{ij}$ is
\begin{equation}\label{the stress tensor}
 \sigma_{ij} =
 \begin{pmatrix}
\sigma_{11}
 & \sigma_{12}
 & \sigma_{13}\\
\sigma_{21}
 & \sigma_{22}
 & \sigma_{23}\\
\sigma_{31}
 & \sigma_{32}
 & \sigma_{33}
 \end{pmatrix}
\end{equation}

For convenience we can introduce the definitions of mean stress
$\sigma_m$ and stress deviator $s_{ij}$ as
\begin{equation}
 \sigma_m
 = \frac{1}{3}(\sigma_{xx} + \sigma_{yy} + \sigma_{zz}),
\end{equation}

\begin{equation}
 s_{ij} =\sigma_{ij} - \sigma_m
 =
 \begin{pmatrix}
\sigma_{11} - \sigma_m
 & \sigma_{12}
 & \sigma_{13}\\
\sigma_{21}
 & \sigma_{22} - \sigma_m
 & \sigma_{23}\\
\sigma_{31}
 & \sigma_{32}
 & \sigma_{33} - \sigma_m
 \end{pmatrix}.
\end{equation}

Now let us turn to study the constitutive relation.

An elastic Hooke solid responds instantaneously with respect to a
external stress. A Newtonian viscous fluid responds to a shear
stress by a steady flow process.

In 19th century, people began to notice that some materials showed
time dependence in their elastic response with respect to external
stress. When materials like pitch, gum rubber, polymeric materials,
hardened cement and even glass were loaded, an instantaneous elastic
deformation was followed by a continuous slow flow or creep.

Now this time-dependent response is called viscoelasticity (refer
to, for instance, \cite{Reiner1960,Christensen1982,Joseph1990}).
Materials exhibits both an instantaneous elastic elasticity and
 creep characteristics is called viscoelastic materials \cite{Christensen1982,Joseph1990}.
 Viscoelastic materials were studied long ago by
Maxwell\cite{Maxwell1868,Christensen1982,Joseph1990}, Kelvin, Voigt,
Boltzamann\cite{Boltzamann1874,Christensen1982,Joseph1990}, etc.

Inspired by these contributors, we propose a visco-elastic
constitutive relation of the aether.

It is natural to speculate that the constitutive relation of the
aether may be a combination of the constitutive relations of the
Hooke-solid and the Newtonian-fluid.

For the Hooke-solid, we have the following the generalized Hooke's
law (refer to, for instance,
\cite{Eringen1975,Fung1977,Wang1982,Landau-Lifshitz1986}),
\begin{equation}\label{the constitutive relation of Hooke-solid 2-1}
 \sigma_{ij} =
2G \varepsilon_{ij} + \lambda \theta \delta_{ij}, \quad
\varepsilon_{ij} = \frac{\sigma_{ij}}{2G} -
\frac{3\nu}{Y}\sigma_{m}\delta_{ij},
\end{equation}
where $\delta_{ij}$ is the Kronecker symbol, $\sigma_{m}$ is the
mean stress, where $Y$ is the Yang modulus, $\nu$ is the Poisson
ratio, $G$ is the shear modulus, $\lambda$ is Lame constant,
$\theta$ is the volume change coefficient. The definition of
$\theta$ is $\theta = \varepsilon_{11} + \varepsilon_{22} +
\varepsilon_{33} = \frac{\partial u}{\partial x} + \frac{\partial
v}{\partial y} + \frac{\partial w}{\partial z}$.

The generalized Hooke's law Eq.(\ref{the constitutive relation of
Hooke-solid 2-1}) can also be written as \cite{Wang1982}
\begin{equation}\label{the constitutive relation of Hooke-solid 2-2}
s_{ij} = 2Ge_{ij},
\end{equation}
where $s_{ij}$ is the stress deviator, $e_{ij}$ is the strain
deviator.

For the Newtonian-fluid, we have the following constitutive relation
\begin{equation}\label{the constitutive relation of Newtonian-fluid}
\frac{d e_{ij}}{d t} = \frac{1}{2\eta}s_{ij},
\end{equation}
where $s_{ij}$ is the stress deviator, $d e_{ij}/d t$ is the strain
rate deviator, $\eta$ is the dynamic viscocity.

Since the aether behaves like the Hooke-solid during very short time
intervals, we therefore differentiate both sides of
 Eq.(\ref{the constitutive relation of Hooke-solid 2-2}) and obtain
\begin{equation}\label{the constitutive relation of Hooke-solid 2}
\frac{d e_{ij}}{d t} = \frac{1}{2G} \frac{d s_{ij}}{d t}.
\end{equation}

A combination of Eq.(\ref{the constitutive relation of
 Hooke-solid 2}) and  Eq.(\ref{the constitutive relation of Newtonian-fluid}) gives
\begin{equation}\label{the constitutive relation of aether 1}
\frac{d e_{ij}}{d t} = \frac{1}{2\eta}s_{ij} + \frac{1}{2G} \frac{d
s_{ij}}{d t}.
\end{equation}

The materials which behave like Eq.(\ref{the constitutive relation
of aether 1}) are called Maxwell-liquid since Maxwell established
such a constitutive relation in 1868 (refer to, for instance,
\cite{Maxwell1868,Reiner1960,Christensen1982,Joseph1990}).

Eq.(\ref{the constitutive relation of aether 1}) is valid only in
the case of infinitesimal deformation because of the presence of the
derivative with respect to time. Oldroyd recognized that we need a
special definition of derivative operation in order to satisfy the
principle of material frame indifference or
objectivity\cite{Oldroyd1950,Christensen1982}. Unfortunately, there
is no unique definition of derivative operation fulfil the the
principle of objectivity presently \cite{Christensen1982}.

As an enlightening example, let us recall the description in
\cite{Reiner1960} about a simple shear experiment. We suppose
\begin{equation}
\frac{d \sigma_{t}}{d t} = \frac{\partial \sigma_{t}}{\partial t},
\quad \frac{d e_{t}}{d t} = \frac{\partial e_{t}}{\partial t},
\end{equation}
where $\sigma_{t}$ is the shear stress, $e_{t}$ is the shear strain.

Therefore, Eq.(\ref{the constitutive relation of aether 1}) becomes
\begin{equation}\label{the constitutive relation of aether 2}
\frac{\partial e_{t}}{\partial t} = \frac{1}{2\eta}\sigma_{t} +
\frac{1}{2G} \frac{\partial \sigma_{t}}{\partial t}.
\end{equation}

Integration of  Eq.(\ref{the constitutive relation of aether 2})
gives
\begin{equation}\label{the constitutive relation of aether 2.5}
\sigma_{t} = e^{-\frac{G}{\eta}t}\left (\sigma_{0} + 2G
\int_{0}^{t}\frac{d e_{t}}{d t} e^{\frac{G}{\eta}} dt \right ).
\end{equation}

If the shear deformation is kept constant, i.e. $\partial
e_{t}/\partial t = 0$, we have
\begin{equation}\label{the constitutive relation of aether 4}
\sigma_{t} = \sigma_{0}e^{-\frac{G}{\eta}t}.
\end{equation}

Eq.(\ref{the constitutive relation of aether 4}) shows that the
shear stresses remain in the Maxwell-liquid and are damped in the
course of time.

We see that $\eta/G$ must have the dimension of time. Now let us
introduce the following definition of Maxwellian relaxation time
$\tau$
\begin{equation}\label{Maxwellian relaxation time 2-1}
\tau = \frac{\eta}{G}.
\end{equation}

Therefore, using Eq.(\ref{Maxwellian relaxation time 2-1}),
Eq.(\ref{the constitutive relation of aether 1}) becomes
\begin{equation}\label{the constitutive relation of aether 3}
 \frac{s_{ij}}{\tau} + \frac{d s_{ij}}{d t} = 2G \frac{d e_{ij}}{d t}.
\end{equation}

Now let us introduce the following hypothesis
\begin{hypothesis}\label{hypothesis of constitutive relation of the aether}
Suppose the constitutive relation of the aether satisfies
Eq.(\ref{the constitutive relation of aether 1}).
\end{hypothesis}

Now we can derive the the equation of momentum conservation based on
the above hypotheses \ref{hypothesis of constitutive relation of the
aether}.

Let $T$ be the characteristic time scale of a observer of the
aether. When the observer's time scale $T$ is the same order of the
period of wave motion of light, the Maxwellian relaxation time
$\tau$ is a relatively a large number. Thus, the first term of
Eq.(\ref{the constitutive relation of aether 3}) may be neglected.
Therefore, the observer concludes
 that the strain and the stress of the aether satisfies the
generalized Hooke's law.

The generalized Hooke's law (\ref{the constitutive relation of
Hooke-solid 2-1}) can also be written as \cite{Fung1977,Wang1982}
\begin{equation}\label{The generalized Hooke's law 2-3}
\begin{aligned}
\sigma_{11} &=&&\lambda \theta+2G\varepsilon_{11}\\
\sigma_{22} &=&&\lambda \theta+2G\varepsilon_{22}\\
\sigma_{33} &=&&\lambda \theta+2G\varepsilon_{33}\\
\sigma_{12} &=&&\sigma_{21} = 2G\varepsilon_{12} = 2G\varepsilon_{21}\\
\sigma_{23} &=&&\sigma_{32} = 2G\varepsilon_{23} = 2G\varepsilon_{32}\\
 \sigma_{31} &=&&\sigma_{13} = 2G\varepsilon_{31}=2G\varepsilon_{13}
\end{aligned}
\end{equation}
where $\lambda = Y\nu/[(1+\nu)(1-2\nu)]$ is Lame constant, $\theta$
is the volume change coefficient. The definition of $\theta$ is
$\theta = \varepsilon_{11} + \varepsilon_{22} + \varepsilon_{33} =
\frac{\partial u}{\partial x} + \frac{\partial v}{\partial y} +
\frac{\partial w}{\partial z}$.

The following relationship are useful
\begin{equation}
G= \frac{Y}{2(1+\nu)},\quad  K= \frac{Y}{3(1-2\nu)},
\end{equation}
where  $K$ is the volume modulus.

It is known that the equations of momentum conservation are (refer
to, for instance,
\cite{Pao-Mow1973,Eringen1975,Fung1977,Wang1982,Pao1983,Landau-Lifshitz1986}),
\begin{eqnarray}
 \frac{\partial \sigma_{11}}{\partial x}+\frac{\partial \sigma_{12}}{\partial y}
 +\frac{\partial \sigma_{13}}{\partial z}+f_{x}=\rho \frac{\partial^2 u}{\partial t^2}, \label{momentum conservation 3-10-1}\\
  \frac{\partial \sigma_{21}}{\partial x}+\frac{\partial \sigma_{22}}{\partial y}
 +\frac{\partial \sigma_{23}}{\partial z}+f_{y}=\rho \frac{\partial^2 v}{\partial t^2}, \label{momentum conservation 3-10-2}\\
  \frac{\partial \sigma_{31}}{\partial x}+\frac{\partial \sigma_{32}}{\partial y}
 +\frac{\partial \sigma_{33}}{\partial z}+f_{z}=\rho \frac{\partial^2 w}{\partial t^2}, \label{momentum conservation 3-10-3}
\end{eqnarray}
where $f_{x}$, $f_{y}$ and $f_{z}$ are the volume force density
exerted on the aether.

The tensor form of the equations (\ref{momentum conservation
3-10-1}-\ref{momentum conservation 3-10-3}) of momentum conservation
can be written as
\begin{equation}
\sigma_{ij,j}+f_{i}=\rho \frac{\partial^2 u_{i}}{\partial t^2}.
\end{equation}

Noticing Eq.(\ref{The generalized Hooke's law 2-3}),
Eqs.(\ref{momentum conservation 3-10-1}-\ref{momentum conservation
3-10-3}) can also be written as
\begin{eqnarray}
 2G\left ( \frac{\partial \varepsilon_{11}}{\partial x}+\frac{\partial \varepsilon_{12}}{\partial y}
 +\frac{\partial \varepsilon_{13}}{\partial z} \right )
 + \lambda \frac{\partial \theta}{\partial x} + f_{x} = \rho \frac{\partial^2 u}{\partial t^2}, \label{momentum conservation 3-20-1}\\
 2G\left (  \frac{\partial \varepsilon_{21}}{\partial x}+\frac{\partial \varepsilon_{22}}{\partial y}
 +\frac{\partial \varepsilon_{23}}{\partial z} \right )
 + \lambda \frac{\partial \theta}{\partial y} + f_{y}=\rho \frac{\partial^2 v}{\partial t^2}, \label{momentum conservation 3-20-2}\\
 2G\left (  \frac{\partial \varepsilon_{31}}{\partial x}+\frac{\partial \varepsilon_{32}}{\partial y}
 +\frac{\partial \varepsilon_{33}}{\partial z} \right )
 + \lambda \frac{\partial \theta}{\partial z} + f_{z}=\rho \frac{\partial^2 w}{\partial t^2}. \label{momentum conservation 3-20-3}
\end{eqnarray}

Using Eq.(\ref{strain-displacements}), Eqs.(\ref{momentum
conservation 3-20-1}-\ref{momentum conservation 3-20-3}) can also be
expressed by means of displacement $\mbox{\upshape\bfseries{u}}$
\begin{equation}\label{momentum conservation 3-30}
\begin{aligned}
 G \nabla^2 u + (G +\lambda ) \frac{\partial }{\partial x}
 \left ( \frac{\partial u}{\partial x} + \frac{\partial v}{\partial y} + \frac{\partial w}{\partial z} \right )
 + f_{x} = \rho \frac{\partial^2 u}{\partial t^2},\\
 G \nabla^2 v + (G +\lambda ) \frac{\partial }{\partial y}
 \left ( \frac{\partial u}{\partial x} + \frac{\partial v}{\partial y} + \frac{\partial w}{\partial z} \right )
 + f_{y} = \rho \frac{\partial^2 v}{\partial t^2},\\
 G \nabla^2 w + (G +\lambda ) \frac{\partial }{\partial z}
 \left ( \frac{\partial u}{\partial x} + \frac{\partial v}{\partial y} + \frac{\partial w}{\partial z} \right )
 + f_{z} = \rho \frac{\partial^2 w}{\partial t^2}.
\end{aligned}
\end{equation}

The vector form of the above equations (\ref{momentum conservation
3-30}) can be written as (refer to, for instance,
\cite{Pao-Mow1973,Eringen1975,Fung1977,Wang1982,Pao1983,Landau-Lifshitz1986}),
\begin{equation}\label{The vector form of the equation of momentum conservation}
 G \nabla^2 \mbox{\upshape\bfseries{u}}
 + (G +\lambda )\nabla (\nabla \cdot \mbox{\upshape\bfseries{u}})
 + \mbox{\upshape\bfseries{f}} = \rho \frac{\partial^2 \mbox{\upshape\bfseries{u}}}{\partial t^2}.
\end{equation}

When there are no body force in the aether, Eqs.(\ref{The vector
form of the equation of momentum conservation}) reduces to
\begin{equation}\label{The vector form of the equation of momentum conservation 2}
 G \nabla^2 \mbox{\upshape\bfseries{u}}
 + (G +\lambda )\nabla (\nabla \cdot \mbox{\upshape\bfseries{u}})
  = \rho \frac{\partial^2 \mbox{\upshape\bfseries{u}}}{\partial t^2}.
\end{equation}

From a theorem of Long \cite{Long1967,Eringen1975}, there exist a
scalar function $\phi$ and a vector function
$\mbox{\upshape\bfseries{R}}$ such that
$\mbox{\upshape\bfseries{u}}$ is represented by
\begin{equation}\label{Stokes-Helmholtz resolution theorem 3-1}
 \mbox{\upshape\bfseries{u}} = \nabla \psi + \nabla \times
 \mbox{\upshape\bfseries{R}}
\end{equation}
and $\phi$ and $\mbox{\upshape\bfseries{R}}$ satisfy the following
wave equations
\begin{eqnarray}
&&\nabla^2 \phi - \frac{1}{c_l} \frac{\partial^2 \phi}{\partial t^2}
= 0,
\label{elastic wave equations 3-1} \\
&&\nabla^2 \mbox{\upshape\bfseries{R}} - \frac{1}{c_t}
\frac{\partial^2 \mbox{\upshape\bfseries{R}}}{\partial t^2} =
0,\label{elastic wave equations equations 3-2}
\end{eqnarray}
where $c_l$ is the velocity of longitudinal waves, $c_t$ is the
velocity of transverse waves. The definitions of these two elastic
wave velocities are (refer to, for instance,
\cite{Pao-Mow1973,Eringen1975,Pao1983,Landau-Lifshitz1986}),
\begin{equation}\label{two elastic wave velocities 3-1}
 c_l = \sqrt{\frac{\lambda + 2G}{\rho}}, \quad
 c_t = \sqrt{\frac{G}{\rho}}.
\end{equation}

$\psi$ and $\mbox{\upshape\bfseries{R}}$ is usually called the
scalar displacement potential and the vector displacement potential
respectively.

\stepcounter{section}
\section{\thesection . \  \label{Point Source and Sink}Definition of Point Source and Sink} If there exists a
velocity field which is continuous and finite at all points of the
space, with the exception of individual isolated points, then these
isolated points are called velocity singularities usually. Point
sources and sinks are examples of velocity singularities.

\begin{definition}\label{source or sink}
Suppose there exists a singularity at point $P_0=(x_0,y_0,z_0)$ in a
continuum.
 If the velocity field of the singularity at point $P=(x,y,z)$ is
\begin{equation}\label{velocity field of source or sink}
\mbox{\upshape\bfseries{v}}(x,y,z,t)=\frac{Q}{4\pi
r^2}\hat{\mbox{\upshape\bfseries{r}}},
\end{equation}
where $r=\sqrt{(x-x_0)^2+(y-y_0)^2+(z-z_0)^2}$,
$\hat{\mbox{\upshape\bfseries{r}}}$
 denotes the unit vector directed outward along the line
from the singularity to the point $P=(x,y,z)$, then we call this
singularity a point source if $Q>0$ or a point sink if $Q<0$. $Q$ is
called the strength of the source or sink.
\end{definition}

Suppose a static point source with strength $Q$ locates at the
origin $(0, 0, 0)$. In order to calculate the volume leaving the
source per unit time, we may enclose the source with an arbitrary
spherical surface $S$ with radius $a$. A calculation shows that
\begin{equation}\label{volume leaving the source 2-10}
\int\hspace{-1.95ex}\int_{S} \hspace{-3.35ex}\bigcirc \
\mbox{\upshape\bfseries{u}}\cdot\mbox{\upshape\bfseries{n}}dS
 = \int\hspace{-1.95ex}\int_{S}
\hspace{-3.35ex}\bigcirc \ \frac{Q}{4\pi
a^2}\hat{\mbox{\upshape\bfseries{r}}}\cdot\mbox{\upshape\bfseries{n}}dS
= Q,
\end{equation}
where $\mbox{\upshape\bfseries{n}}$ denotes the unit vector directed
outward along the line from the origin of the coordinates to the
field point$(x,y,z)$. Equation (\ref{volume leaving the source
2-10}) shows that the strength $Q$ of a source or sink evaluates the
volume of the fluid leaving or entering a control surface per unit
time.

For the case of continuously distributed point sources or sinks, it
is useful to introduce the following definition of volume density
$\rho_s$ of point sources or sinks,
\begin{equation}\label{definition of the volume density of continuously distributed sink}
\rho_s=\lim_{\triangle V \rightarrow 0}\frac{\triangle Q}{\triangle
V},
\end{equation}
where $\triangle V$ is a small volume, $\triangle Q$ is the sum of
strengthes of all the point sources or sinks in the volume
$\triangle V$.

\stepcounter{section}
\section{\thesection . \  \label{Point Source and Sink Model of Electric Charges}A Point Source and Sink Model of
Electric Charges} The purpose of this section is to propose a point
source and sink model of electric charges.

Let $T$ be the characteristic time of a observer of a electric
charge in the aether. We may suppose that the observer's time scale
$T$ is very large compares to the Maxwellian relaxation time $\tau$.
So the Maxwellian relaxation time $\tau$ is a relatively a small
number and the stress deviator $s_{ij}$ changes very slowly. Thus,
the second term in the left of Eq.(\ref{the constitutive relation of
aether 3}) may be neglected. According to this observer,
 the constitutive relation of the aether may be written as
\begin{equation}\label{the constitutive relation of aether 4-3}
 s_{ij} = 2 \eta \frac{d e_{ij}}{d t}.
\end{equation}

Therefore, the observer concludes
 that the aether behaves like the Newtonian-fluid under his time scale.

In order to compare fluid motion with electric fields, Maxwell
introduced an analogy between sources or sinks and electric charges
\cite{Whittaker1953}. Inspired by Maxwell, we may introduce the
following
\begin{hypothesis}\label{electric charges hypothesis}
Suppose all the electric charges in the universe are sources or
sinks in the aether. We define a source as a negative electric
charge. We define a sink as a positive electric charge. The electric
charge quantity $q_{e}$ of a electric charge is defined as
\begin{equation}\label{definition of electric charge quantity 5-10}
q_{e} = - k_{Q}\rho Q,
\end{equation}
where $\rho$ is the density of the aether, $Q$ is called the
strength of the source or sink, $k_{Q}$ is a positive dimensionless
constant.
\end{hypothesis}

A calculation shows that the mass $m$ of a electric charge is
changing with time as
\begin{equation}\label{the mass is changing}
\frac{dm}{dt} = -\rho Q = \frac{q_{e}}{k_{Q}},
\end{equation}
where $q_{e}$ is the electric charge quantity of the electric
charge.

We may introduce a hypothesis that the masses of electric charges
are changing so slowly relative to the time scaler of human beings
that they can be treated as constants approximately.

For the case of continuously distributed electric charges, it is
useful to introduce the following definition of volume density
$\rho_e$ of electric charges
\begin{equation}\label{continuously distributed electric charges}
\rho_e = \lim_{\triangle V \rightarrow 0}\frac{\triangle
q_{e}}{\triangle V}
\end{equation}
where $\triangle V$ is a small volume, $\triangle q_{e}$ is the sum
of strengthes of all the electric charges in the volume $\triangle
V$.

From Eq.(\ref{definition of the volume density of continuously
distributed sink}), Eq.(\ref{definition of electric charge quantity
5-10}) and Eq.(\ref{continuously distributed electric charges}), we
have
\begin{equation}\label{relation between charge density and sink density}
\rho_e = - k_{Q} \rho \rho_{s}
\end{equation}

\stepcounter{section}
\section{\thesection . \  \label{Maxwell's Equations in vacuum}Derivation of the Maxwell's Equations in vacuum}
The purpose of this section is to derive the Maxwell's equations
based on the above visco-elastic continuum model of vacuum and the
singularity model of electric charges.

Now let us to derive the continuity equation of the aether from mass
conservation. Consider an arbitrary volume $V$ bounded by a closed
surface $S$ fixed in space. Suppose there are electric charges
continuously distributed in the volume $V$. The total mass in volume
$V$ is
\begin{equation}
M=\int\hspace{-1.5ex}\int\hspace{-1.5ex}\int_{V} \rho dV,
\end{equation}
where $\rho$ is the density of the aether.

The rate of increase of the total mass in volume $V$ is
\begin{equation}\label{The rate of increase of the total mass}
 \frac{\partial M}{\partial t}=\frac{\partial }{\partial t}\int\hspace{-1.5ex}\int\hspace{-1.5ex}\int_{V}\rho
 dV.
\end{equation}

Using the Ostrogradsky--Gauss theorem (refer to, for instance,
\cite{Lamb1932,Kochin1964,Yih1969,Wu1982a,Landau-Lifshitz1987,Faber1995,Currie2003}),
the rate of mass outflow through the surface $S$ is
\begin{equation}\label{The rate of mass outflow 5-10}
 \int\hspace{-1.9ex}\int_{S}\hspace{-3.27ex}\bigcirc\rho (\mbox{\bfseries{u}} \cdot \mbox{\bfseries{n}}
 )dS = \int\hspace{-1.5ex}\int\hspace{-1.5ex}\int_{V} \nabla \cdot (\rho \mbox{\bfseries{u}})dV,
\end{equation}
where $\mbox{\bfseries{v}}$ is the velocity field of the aether.

The definition of the velocity field $\mbox{\bfseries{v}}$ is
\begin{equation}\label{velocity}
v_i = \frac{\partial u_i}{\partial t}, \quad or \quad
\mbox{\bfseries{v}} = \frac{\partial \mbox{\bfseries{u}}}{\partial
t}.
\end{equation}

Using Eq.(\ref{relation between charge density and sink density}),
the rate of mass created inside the volume $V$ is
\begin{equation}\label{the rate of mass created}
\int\hspace{-1.5ex}\int\hspace{-1.5ex}\int_{V} \rho\rho_{s}dV =
\int\hspace{-1.5ex}\int\hspace{-1.5ex}\int_{V}
-\frac{\rho_{e}}{k_{Q}}dV.
\end{equation}

Now according to the principle of mass conservation, and making use
of Eq.(\ref{The rate of increase of the total mass}), Eq.(\ref{The
rate of mass outflow 5-10})
 and  Eq.(\ref{the rate of mass created}), we have
\begin{equation}\label{equation of mass conservation 5-10}
 \frac{\partial }{\partial t}\int\hspace{-1.5ex}\int\hspace{-1.5ex}\int_{V} \rho dV
 = \int\hspace{-1.5ex}\int\hspace{-1.5ex}\int_{V} -\frac{\rho_{e}}{k_{Q}}dV - \int\hspace{-1.5ex}\int\hspace{-1.5ex}\int_{V} \nabla \cdot (\rho \mbox{\bfseries{v}})dV
\end{equation}

Since the volume $V$ is arbitrary, from Eq.(\ref{equation of mass
conservation 5-10}) we have
\begin{equation}\label{equation of mass conservation 5-20}
 \frac{\partial \rho}{\partial t} + \nabla \cdot (\rho \mbox{\bfseries{v}}) = -\frac{\rho_{e}}{k_{Q}}.
\end{equation}

According to Hypothesis \ref{homogeneous hypothesis 3-10}, the
aether is homogeneous, that is
\begin{math}
\frac{\partial\rho}{\partial x}=\frac{\partial\rho}{\partial y}
 =\frac{\partial\rho}{\partial z}=\frac{\partial\rho}{\partial
 t}=0.
\end{math}
Thus, Eq.(\ref{equation of mass conservation 5-20}) becomes
\begin{equation}\label{equation of mass conservation 5-30}
 \nabla \cdot  \mbox{\bfseries{v}} = -\frac{\rho_{e}}{k_{Q}\rho}.
\end{equation}

According to Hypothesis \ref{electric charges hypothesis} and
Eq.(\ref{the mass is changing}), the masses of positive electric
charges are changing since the strength of a sink evaluates the
volume of the aether entering the sink per unit time. Thus, the
momentum of a volume element $\triangle V$ of the aether containing
continuously distributed electric charges moving with an average
speed $\mbox{\upshape\bfseries{v}}_{e}$ is changing. The increased
momentum $\triangle\mbox{\upshape\bfseries{P}}$ of the volume
element $\triangle V$ during a time interval $\triangle t$ is the
decreased momentum of continuously distributed electric charges
contained in the volume element $\triangle V$ during a time interval
$\triangle t$, that is,
\begin{equation}
\triangle\mbox{\upshape\bfseries{P}} = \rho (\rho_{s}\triangle V
\triangle t) \mbox{\upshape\bfseries{v}}_{e}= -
\frac{\rho_{e}}{k_{Q}}\triangle V \triangle t
\mbox{\upshape\bfseries{v}}_{e}
\end{equation}

Therefore, the equation of momentum
 conservation Eq.(\ref{The vector form of the equation of momentum conservation})
 of the aether should be changed as
\begin{equation}\label{reduced equation of momentum conservation 5-1}
 G \nabla^2 \mbox{\upshape\bfseries{u}}
 + (G +\lambda )\nabla (\nabla \cdot \mbox{\upshape\bfseries{u}})
 + \mbox{\upshape\bfseries{f}}
 = \rho \frac{\partial^2 \mbox{\upshape\bfseries{u}}}{\partial t^2}
 - \frac{\rho_{e}\mbox{\upshape\bfseries{v}}_{e}}{k_{Q}}.
\end{equation}

In order to simplify the Eq.(\ref{reduced equation of momentum
conservation 5-1}), we may introduce an additional hypothesis as
\begin{hypothesis}\label{incompressible hypothesis}
We suppose that the aether is incompressible, that is $\theta =0$.
\end{hypothesis}

 From  Hypothesis \ref{incompressible hypothesis}, we have
\begin{equation}
 \nabla \cdot \mbox{\upshape\bfseries{u}}
 = \frac{\partial u}{\partial x}+\frac{\partial v}{\partial y}+\frac{\partial w}{\partial z}
 = \theta
 = 0
\end{equation}

 Therefore, the vector form of the equation of momentum
 conservation Eq.(\ref{reduced equation of momentum
conservation 5-1})
 reduces to the following form
\begin{equation}\label{reduced equation of momentum conservation 5-2}
 G \nabla^2 \mbox{\upshape\bfseries{u}}+\mbox{\upshape\bfseries{f}}
 = \rho \frac{\partial^2 \mbox{\upshape\bfseries{u}}}{\partial t^2} - \frac{\rho_{e}\mbox{\upshape\bfseries{v}}_{e}}{k_{Q}}.
\end{equation}

According to the Stokes-Helmholtz resolution theorem (refer to, for
instance, \cite{Pao-Mow1973,Eringen1975}), which states that every
sufficiently smooth vector field may be decomposed into irrotational
and solenoidal parts, there exist a scalar function $\phi$ and a
vector function $\phi$ and a $\mbox{\upshape\bfseries{R}}$ such that
$\mbox{\upshape\bfseries{u}}$ is represented by
\begin{equation}\label{Stokes-Helmholtz resolution theorem}
 \mbox{\upshape\bfseries{u}} = \nabla \psi + \nabla \times
 \mbox{\upshape\bfseries{R}}.
\end{equation}

Now let us introduce the definitions
\begin{equation}\label{definition of electromagnetic potentials}
 \nabla\phi = k_{E}\frac{\partial}{\partial t}(\nabla \psi),
 \quad \mbox{\upshape\bfseries{A}} = k_{E}\nabla \times
 \mbox{\upshape\bfseries{R}},
\end{equation}
\begin{equation}\label{definition of electric field intensity and magnetic induction}
 \mbox{\upshape\bfseries{E}} =
 -k_{E}\frac{\partial \mbox{\upshape\bfseries{u}}}{\partial t}, \quad
 \mbox{\upshape\bfseries{B}} = k_{E} \nabla \times
 \mbox{\upshape\bfseries{u}},
\end{equation}
 where $\phi$ is the scalar electromagnetic potential, $\mbox{\upshape\bfseries{A}}$ is the
vector electromagnetic potential, $\mbox{\upshape\bfseries{E}}$ is
the electric field intensity, $\mbox{\upshape\bfseries{B}}$ is the
magnetic induction, $k_{E}$ is a positive dimensionless constant.

From Eq.(\ref{Stokes-Helmholtz resolution theorem}),
Eq.(\ref{definition of electromagnetic potentials}) and
Eq.(\ref{definition of electric field intensity and magnetic
induction}), we have
\begin{equation}\label{equations of electric field intensity and magnetic induction 6-10}
\mbox{\upshape\bfseries{E}} =
 -\nabla \phi - \frac{\partial \mbox{\upshape\bfseries{A}}}{\partial
 t}, \quad
 \mbox{\upshape\bfseries{B}} = \nabla \times \mbox{\upshape\bfseries{A}}
\end{equation}
and
\begin{eqnarray}
 \nabla \times\mbox{\upshape\bfseries{E}} &=&
 -\frac{\partial \mbox{\upshape\bfseries{B}}}{\partial t}, \label{equations of electric field intensity and magnetic induction 6-11}\\
 \nabla \cdot \mbox{\upshape\bfseries{B}}&=& 0.\label{equations of electric field intensity and magnetic induction 6-12}
 \end{eqnarray}

Based on Eq.(\ref{definition of electromagnetic potentials}) and
noticing
\begin{eqnarray}
 \nabla^2 \mbox{\upshape\bfseries{u}} &= \nabla(\nabla \cdot
 \mbox{\upshape\bfseries{u}}) - \nabla \times (\nabla \times
 \mbox{\upshape\bfseries{u}}), \label{identity of a vector field 6-11}\\
 \nabla^2 \mbox{\upshape\bfseries{A}} &= \nabla(\nabla \cdot
 \mbox{\upshape\bfseries{A}}) - \nabla \times (\nabla \times
 \mbox{\upshape\bfseries{A}}),\label{identity of a vector field 6-12}
\end{eqnarray}
and \begin{math} \nabla \cdot \mbox{\upshape\bfseries{u}}=0
\end{math}, \begin{math}
\nabla \cdot \mbox{\upshape\bfseries{A}}=0
\end{math}, we have
\begin{equation}\label{electromagnetic potential and displacement vector 6-10}
 k_{E}\nabla^2 \mbox{\upshape\bfseries{u}} = \nabla^2
 \mbox{\upshape\bfseries{A}}.
\end{equation}

Therefore, using Eq.(\ref{electromagnetic potential and displacement
vector 6-10}), Eq.(\ref{reduced equation of momentum conservation
5-2}) becomes
\begin{equation}\label{electromagnetic form of equation of momentum conservation 1}
  \frac{G}{k_{E}} \nabla^2 \mbox{\upshape\bfseries{A}}+\mbox{\upshape\bfseries{f}}
  = \rho \frac{\partial^2 \mbox{\upshape\bfseries{u}}}{\partial t^2} - \frac{\rho_{e}\mbox{\upshape\bfseries{v}}_{e}}{k_{Q}}.
\end{equation}

Using Eq.(\ref{identity of a vector field 6-12}),
 Eq.(\ref{electromagnetic form of equation of momentum conservation 1}) becomes
\begin{equation}\label{electromagnetic form of equation of momentum conservation 2}
 -\frac{G}{k_{E}} \nabla \times (\nabla \times \mbox{\upshape\bfseries{A}}) +\mbox{\upshape\bfseries{f}}
 = \rho \frac{\partial^2 \mbox{\upshape\bfseries{u}}}{\partial
 t^2} - \frac{\rho_{e}\mbox{\upshape\bfseries{v}}_{e}}{k_{Q}}.
\end{equation}

Now using Eq.(\ref{equations of electric field intensity and
magnetic induction 6-10}),
 Eq.(\ref{electromagnetic form of equation of momentum conservation 2}) becomes
\begin{equation}\label{electromagnetic form of equation of momentum conservation 3}
 -\frac{G}{k_{E}} \nabla \times \mbox{\upshape\bfseries{B}} +\mbox{\upshape\bfseries{f}}
  = -\frac{\rho}{k_{E}} \frac{\partial \mbox{\upshape\bfseries{E}}}{\partial
 t} - \frac{\rho_{e}\mbox{\upshape\bfseries{v}}_{e}}{k_{Q}}.
\end{equation}

It is natural to speculate that there are no other body forces or
surface forces exerted on the aether. Thus, we have
$\mbox{\upshape\bfseries{f}}=0$. Therefore, Eq.(\ref{electromagnetic
form of equation of momentum conservation 3}) becomes
\begin{equation}\label{electromagnetic form of equation of momentum conservation 4}
 \frac{k_{Q}G}{k_{E}} \nabla \times \mbox{\upshape\bfseries{B}} = \frac{k_{Q}\rho}{k_{E}} \frac{\partial \mbox{\upshape\bfseries{E}}}{\partial
 t} +  \rho_{e} \mbox{\upshape\bfseries{v}}_{e}.
\end{equation}

Now let us introduce the following definitions
\begin{equation}\label{3 defitions}
 \mbox{\upshape\bfseries{j}} = \rho_{e}
 \mbox{\upshape\bfseries{v}}_{e}, \quad \epsilon_{0} = \frac{k_{Q}\rho}{k_{E}}, \quad
  \frac{1}{\mu_{0}} = \frac{k_{Q}G}{k_{E}}.
\end{equation}

Therefore, Eq.(\ref{electromagnetic form of equation of momentum
conservation 4}) becomes
\begin{equation}\label{electromagnetic form of equation of momentum conservation 5}
 \frac{1}{\mu_{0}}\nabla \times\mbox{\upshape\bfseries{B}}
 =\mbox{\upshape\bfseries{j}}+\epsilon_{0}\frac{\partial \mbox{\upshape\bfseries{E}}}{\partial t}
\end{equation}

Noticing Eq.(\ref{definition of electric field intensity and
magnetic induction}) and Eq.(\ref{3 defitions}), Eq.(\ref{equation
of mass conservation 5-30}) becomes
\begin{equation}\label{equation of mass conservation 3}
 \nabla \cdot \mbox{\upshape\bfseries{E}} = \frac{\rho_{e}}{\epsilon_{0}}.
\end{equation}

Now we see that Eq.(\ref{equations of electric field intensity and
magnetic induction 6-11}), Eq.(\ref{equations of electric field
intensity and magnetic induction 6-12}), Eq.(\ref{electromagnetic
form of equation of momentum conservation 5}) and Eq.(\ref{equation
of mass conservation 3}) coincide with the Maxwell's equations
(\ref{The Maxwell's equations 1-10}-\ref{The Maxwell's equations
1-40}).

\stepcounter{section}
\section{\thesection . \  \label{Electromagnetic Wave}Mechanical Interpretation of the Electromagnetic Wave}
It is known that from the Maxwell's equations (\ref{The Maxwell's
equations 1-10}-\ref{The Maxwell's equations 1-40}), we can obtain
the following equations (refer to, for instance, \cite{Jackson1963})
\begin{eqnarray}
&&\nabla^2 \mbox{\upshape\bfseries{E}} -
\frac{1}{\mu_{0}\epsilon_{0}} \frac{\partial^2
\mbox{\upshape\bfseries{E}}}{\partial t^2} = \frac{1}{\epsilon_{0}}
\nabla \rho_e + \mu_{0}\frac{\partial \mbox{\upshape\bfseries{j}}
}{\partial t},\label{Electromagnetic Wave equations 7-1}\\
&&\nabla^2 \mbox{\upshape\bfseries{H}} -
\frac{1}{\mu_{0}\epsilon_{0}} \frac{\partial^2
\mbox{\upshape\bfseries{H}}}{\partial t^2} = - \frac{1}{\mu_{0}}
\nabla \times \mbox{\upshape\bfseries{j}}.\label{Electromagnetic
Wave equations 7-2}
\end{eqnarray}

Eq.(\ref{Electromagnetic Wave equations 7-1}) and
Eq.(\ref{Electromagnetic Wave equations 7-2}) are the
electromagnetic wave equations with sources in the aether. In the
source free region where $\rho_e = 0$ and
$\mbox{\upshape\bfseries{j}} = 0$,  this equations reduce to the
following equations
\begin{eqnarray}
&&\nabla^2 \mbox{\upshape\bfseries{E}} -
\frac{1}{\mu_{0}\epsilon_{0}} \frac{\partial^2
\mbox{\upshape\bfseries{E}}}{\partial t^2} = 0,
\label{Electromagnetic Wave equations 7-3} \\
&&\nabla^2 \mbox{\upshape\bfseries{H}} -
\frac{1}{\mu_{0}\epsilon_{0}} \frac{\partial^2
\mbox{\upshape\bfseries{H}}}{\partial t^2} =
0.\label{Electromagnetic Wave equations 7-4}
\end{eqnarray}
Eq.(\ref{Electromagnetic Wave equations 7-3}) and
Eq.(\ref{Electromagnetic Wave equations 7-4}) are the
electromagnetic wave equations without sources in the aether.

From Eq.(\ref{Electromagnetic Wave equations 7-3}),
Eq.(\ref{Electromagnetic Wave equations 7-4}) and Eq.(\ref{3
defitions}), we see that the velocity $c_0$ of electromagnetic waves
in vacuum is
\begin{equation}\label{velocity of electromagnetic waves 7-1}
 c_0 = \frac{1}{\sqrt{\mu_{0}\epsilon_{0}}} = \sqrt{\frac{G}{\rho}}.
\end{equation}

From Eq.(\ref{two elastic wave velocities 3-1}) and
Eq.(\ref{velocity of electromagnetic waves 7-1}), we see that the
velocity $c_0$ of electromagnetic waves in the vacuum is the same as
the velocity $c_t$ of the transverse elastic waves in the aether.

Now we may regard electromagnetic waves in the vacuum as the
transverse waves in the aether. This idea was first introduced by
Fresnel in 1821 \cite{Whittaker1953}.

\stepcounter{section}
\section{\thesection . \  \label{Conclusion}Conclusion}
It is an old idea that the universe may be filled with a kind of
continuously distributed material which may be called aether.
Following Stokes, we propose a visco-elastic constitutive relation
of the aether. Following Maxwell, we propose a fluidic source and
sink model model of electric charges. Thus, the Maxwell's equations
in vacuum are derived by methods of continuum mechanics based on
this continuum mechanical model of vacuum and the singularity model
of electric charges.

\stepcounter{section}
\section{\thesection . \  \label{Discussion}Discussion}
There exists some interesting theoretical, experimental and applied
problems in the fields of continuum mechanics, the classical
electrodynamics, quantum electrodynamics and other related fields
involving this theory of electromagnetic phenomena. It is an
interesting task to generalize this theory of electromagnetic
phenomena in the static aether to the case of electromagnetic
phenomena of moving bodies.



\begin{thebibliography}{62}
\expandafter\ifx\csname
natexlab\endcsname\relax\def\natexlab#1{#1}\fi
\expandafter\ifx\csname bibnamefont\endcsname\relax
  \def\bibnamefont#1{#1}\fi
\expandafter\ifx\csname bibfnamefont\endcsname\relax
  \def\bibfnamefont#1{#1}\fi
\expandafter\ifx\csname citenamefont\endcsname\relax
  \def\citenamefont#1{#1}\fi
\expandafter\ifx\csname url\endcsname\relax
  \def\url#1{\texttt{#1}}\fi
\expandafter\ifx\csname urlprefix\endcsname\relax\def\urlprefix{URL
}\fi \providecommand{\bibinfo}[2]{#2}
\providecommand{\eprint}[2][]{\url{#2}}

\bibitem[{\citenamefont{Jackson}(1963)}]{Jackson1963}
\bibinfo{author}{\bibfnamefont{J.~D.} \bibnamefont{Jackson}},
  \emph{\bibinfo{title}{Classical Electrodynamics}} (\bibinfo{publisher}{Wiley,
  New York}, \bibinfo{year}{1963}).

\bibitem[{\citenamefont{Landau and Lifshitz}(1958)}]{Landau-Lifshitz1958}
\bibinfo{author}{\bibfnamefont{L.~D.} \bibnamefont{Landau}} \bibnamefont{and}
  \bibinfo{author}{\bibnamefont{Lifshitz}},
  \emph{\bibinfo{title}{Non-relativistic Theory Quantum Mechanics, translated
  from the Russian by J.B. Sykes and J.S. Bell.}}
  (\bibinfo{publisher}{Pergamon, London}, \bibinfo{year}{1958}).

\bibitem[{\citenamefont{Aharonov and Bohm}(1959)}]{Aharonov-Bohm1959}
\bibinfo{author}{\bibfnamefont{Y.~A.} \bibnamefont{Aharonov}} \bibnamefont{and}
  \bibinfo{author}{\bibfnamefont{D.}~\bibnamefont{Bohm}},
  \bibinfo{journal}{Phys. Rev} \textbf{\bibinfo{volume}{115}},
  \bibinfo{pages}{485} (\bibinfo{year}{1959}).

\bibitem[{\citenamefont{Chambers}(1960)}]{Chambers1960}
\bibinfo{author}{\bibfnamefont{R.~G.} \bibnamefont{Chambers}},
  \bibinfo{journal}{Phys. Rev. Lett.} \textbf{\bibinfo{volume}{5}},
  \bibinfo{pages}{3} (\bibinfo{year}{1960}).

\bibitem[{\citenamefont{Landau and Lifshitz}(1975)}]{Landau-Lifshitz1975}
\bibinfo{author}{\bibfnamefont{L.~D.} \bibnamefont{Landau}} \bibnamefont{and}
  \bibinfo{author}{\bibnamefont{Lifshitz}}, \emph{\bibinfo{title}{The classical
  theory of fields, translated from the Russian by Morton Hamermesh}}
  (\bibinfo{publisher}{Pergamon Press, New York}, \bibinfo{year}{1975}).

\bibitem[{\citenamefont{Whitney}(1988)}]{Whitney1988}
\bibinfo{author}{\bibfnamefont{C.~K.} \bibnamefont{Whitney}},
  \bibinfo{journal}{Hadronic J.} \textbf{\bibinfo{volume}{11}},
  \bibinfo{pages}{257} (\bibinfo{year}{1988}).

\bibitem[{\citenamefont{Chubykalo and
  Smirnov-Rueda}(1996)}]{Chubykalo-Smirnov-Rueda1996}
\bibinfo{author}{\bibfnamefont{A.~E.} \bibnamefont{Chubykalo}}
  \bibnamefont{and}
  \bibinfo{author}{\bibfnamefont{R.}~\bibnamefont{Smirnov-Rueda}},
  \bibinfo{journal}{Phys. Rev. E} \textbf{\bibinfo{volume}{53}},
  \bibinfo{pages}{5373} (\bibinfo{year}{1996}).

\bibitem[{\citenamefont{Dirac}(1978)}]{Dirac1978}
\bibinfo{author}{\bibfnamefont{P.~A.~M.} \bibnamefont{Dirac}},
  \emph{\bibinfo{title}{Direction in Physics}} (\bibinfo{publisher}{Wiley, New
  York}, \bibinfo{year}{1978}).

\bibitem[{\citenamefont{Lamoreaux}(2005)}]{Lamoreaux2005}
\bibinfo{author}{\bibfnamefont{S.~K.} \bibnamefont{Lamoreaux}},
  \bibinfo{journal}{Rep. Prog. Phy.} \textbf{\bibinfo{volume}{68}},
  \bibinfo{pages}{201} (\bibinfo{year}{2005}).

\bibitem[{\citenamefont{Intravaia and
  Lambrecht}(2005)}]{Intravaia-Lambrecht2005}
\bibinfo{author}{\bibfnamefont{F.}~\bibnamefont{Intravaia}} \bibnamefont{and}
  \bibinfo{author}{\bibfnamefont{A.}~\bibnamefont{Lambrecht}},
  \bibinfo{journal}{Phys. Rev. Lett.} \textbf{\bibinfo{volume}{94}},
  \bibinfo{pages}{110404} (\bibinfo{year}{2005}).

\bibitem[{\citenamefont{Guo and Zhao}(2004)}]{Guo-Zhao2004}
\bibinfo{author}{\bibfnamefont{J.~G.} \bibnamefont{Guo}} \bibnamefont{and}
  \bibinfo{author}{\bibfnamefont{Y.~P.} \bibnamefont{Zhao}},
  \bibinfo{journal}{Journal of Microelectromechanical Systems}
  \textbf{\bibinfo{volume}{13}}, \bibinfo{pages}{1027} (\bibinfo{year}{2004}).

\bibitem[{\citenamefont{Davies}(2005)}]{Davies2005}
\bibinfo{author}{\bibfnamefont{P.~C.~W.} \bibnamefont{Davies}},
  \bibinfo{journal}{J. Opt. B: Quantum Semiclass. Opt.}
  \textbf{\bibinfo{volume}{7}}, \bibinfo{pages}{S40} (\bibinfo{year}{2005}).

\bibitem[{\citenamefont{Truesdell}(1966)}]{Truesdell1966}
\bibinfo{author}{\bibfnamefont{C.}~\bibnamefont{Truesdell}},
  \emph{\bibinfo{title}{The Elements of Continuum Mechanics}}
  (\bibinfo{publisher}{Springer-Verlag, New York}, \bibinfo{year}{1966}).

\bibitem[{\citenamefont{Fung}(1977)}]{Fung1977}
\bibinfo{author}{\bibfnamefont{Y.~C.} \bibnamefont{Fung}},
  \emph{\bibinfo{title}{A First Course in Continuum Mechanics}}
  (\bibinfo{publisher}{Prentice-Hall, London}, \bibinfo{year}{1977}).

\bibitem[{\citenamefont{Eringen}(1980)}]{Eringen1980}
\bibinfo{author}{\bibfnamefont{A.~C.} \bibnamefont{Eringen}},
  \emph{\bibinfo{title}{The Elements of Continuum Mechanics}}
  (\bibinfo{publisher}{Robert E. Krieger Pub. Co., Huntington},
  \bibinfo{year}{1980}).

\bibitem[{\citenamefont{Landau and Lifshitz}(1987)}]{Landau-Lifshitz1987}
\bibinfo{author}{\bibfnamefont{L.~D.} \bibnamefont{Landau}} \bibnamefont{and}
  \bibinfo{author}{\bibnamefont{Lifshitz}}, \emph{\bibinfo{title}{Fluid
  Mechanics, translated from the Russian by J.B. Sykes and W.H. Reid.}}
  (\bibinfo{publisher}{Pergamon, New York}, \bibinfo{year}{1987}).

\bibitem[{\citenamefont{Whittaker}(1953)}]{Whittaker1953}
\bibinfo{author}{\bibfnamefont{E.}~\bibnamefont{Whittaker}},
  \emph{\bibinfo{title}{A History of the Theories of Aether and Electricity}}
  (\bibinfo{publisher}{Thomas Nelson and Sons Ltd., London},
  \bibinfo{year}{1953}).

\bibitem[{\citenamefont{Ellis}(2003)}]{Ellis2003}
\bibinfo{author}{\bibfnamefont{J.}~\bibnamefont{Ellis}},
  \bibinfo{journal}{Philos. Trans. R. Soc. London, A}
  \textbf{\bibinfo{volume}{361}}, \bibinfo{pages}{2607} (\bibinfo{year}{2003}).

\bibitem[{\citenamefont{Linder}(2004)}]{Linder2004}
\bibinfo{author}{\bibfnamefont{E.~V.} \bibnamefont{Linder}},
  \bibinfo{journal}{Phys. Rev. D} \textbf{\bibinfo{volume}{70}},
  \bibinfo{pages}{023511} (\bibinfo{year}{2004}).

\bibitem[{\citenamefont{Bernardeau}(2003)}]{Bernardeau2003}
\bibinfo{author}{\bibfnamefont{F.}~\bibnamefont{Bernardeau}},
  \bibinfo{journal}{Rep. Prog. Phys.} \textbf{\bibinfo{volume}{66}},
  \bibinfo{pages}{691} (\bibinfo{year}{2003}).

\bibitem[{\citenamefont{Bergstrom et~al.}(2005)\citenamefont{Bergstrom,
  Bringmann, and M.~Eriksson}}]{Bergstrom2005}
\bibinfo{author}{\bibfnamefont{L.}~\bibnamefont{Bergstrom}},
  \bibinfo{author}{\bibfnamefont{T.}~\bibnamefont{Bringmann}},
  \bibnamefont{and} \bibinfo{author}{\bibfnamefont{{\itshape{ et
  al.}}.}~\bibnamefont{M.~Eriksson}}, \bibinfo{journal}{Phys. Rev. Lett.}
  \textbf{\bibinfo{volume}{94}}, \bibinfo{pages}{131301}
  (\bibinfo{year}{2005}).

\bibitem[{\citenamefont{Beacom et~al.}(2005)\citenamefont{Beacom, Bell, and
  Bertone}}]{Beacom2005}
\bibinfo{author}{\bibfnamefont{J.~F.} \bibnamefont{Beacom}},
  \bibinfo{author}{\bibfnamefont{N.~F.} \bibnamefont{Bell}}, \bibnamefont{and}
  \bibinfo{author}{\bibfnamefont{G.}~\bibnamefont{Bertone}},
  \bibinfo{journal}{Phys. Rev. Lett.} \textbf{\bibinfo{volume}{94}},
  \bibinfo{pages}{171301} (\bibinfo{year}{2005}).

\bibitem[{\citenamefont{Akerib et~al.}(2006)\citenamefont{Akerib, Attisha, and
  C.~N.~Bailey}}]{Akerib2006}
\bibinfo{author}{\bibfnamefont{D.~S.} \bibnamefont{Akerib}},
  \bibinfo{author}{\bibfnamefont{M.~J.} \bibnamefont{Attisha}},
  \bibnamefont{and} \bibinfo{author}{\bibfnamefont{{\itshape{ et
  al.}}.}~\bibnamefont{C.~N.~Bailey}}, \bibinfo{journal}{Phys. Rev. Lett.}
  \textbf{\bibinfo{volume}{96}}, \bibinfo{pages}{011302}
  (\bibinfo{year}{2006}).

\bibitem[{\citenamefont{Feng et~al.}(2006)\citenamefont{Feng, Su, and
  Takayama}}]{Feng2006}
\bibinfo{author}{\bibfnamefont{J.~L.} \bibnamefont{Feng}},
  \bibinfo{author}{\bibfnamefont{S.~F.} \bibnamefont{Su}}, \bibnamefont{and}
  \bibinfo{author}{\bibfnamefont{F.}~\bibnamefont{Takayama}},
  \bibinfo{journal}{Phys. Rev. Lett.} \textbf{\bibinfo{volume}{96}},
  \bibinfo{pages}{151802} (\bibinfo{year}{2006}).

\bibitem[{\citenamefont{P.~Fayet}(2006)}]{Fayet2006}
\bibinfo{author}{\bibfnamefont{G.~S.} \bibnamefont{P.~Fayet},
  \bibfnamefont{D.~Hooper}}, \bibinfo{journal}{Phys. Rev. Lett.}
  \textbf{\bibinfo{volume}{96}}, \bibinfo{pages}{211302}
  (\bibinfo{year}{2006}).

\bibitem[{\citenamefont{Carena et~al.}(2006)\citenamefont{Carena, Hooper, and
  Skands}}]{Carena2006}
\bibinfo{author}{\bibfnamefont{M.}~\bibnamefont{Carena}},
  \bibinfo{author}{\bibfnamefont{D.}~\bibnamefont{Hooper}}, \bibnamefont{and}
  \bibinfo{author}{\bibfnamefont{P.}~\bibnamefont{Skands}},
  \bibinfo{journal}{Phys. Rev. Lett.} \textbf{\bibinfo{volume}{97}},
  \bibinfo{pages}{051801} (\bibinfo{year}{2006}).

\bibitem[{\citenamefont{Einstein}(1905)}]{Einstein1905}
\bibinfo{author}{\bibfnamefont{A.}~\bibnamefont{Einstein}},
  \bibinfo{journal}{Ann. Phys.} \textbf{\bibinfo{volume}{17}},
  \bibinfo{pages}{891} (\bibinfo{year}{1905}).

\bibitem[{\citenamefont{Golebiewska-Lasta}(1979)}]{Golebiewska-Lasta1979}
\bibinfo{author}{\bibfnamefont{A.~A.} \bibnamefont{Golebiewska-Lasta}},
  \bibinfo{journal}{Int. J. Engng. Sic.} \textbf{\bibinfo{volume}{17}},
  \bibinfo{pages}{441} (\bibinfo{year}{1979}).

\bibitem[{\citenamefont{Barut}(1988)}]{Barut1988}
\bibinfo{author}{\bibfnamefont{A.}~\bibnamefont{Barut}},
  \bibinfo{journal}{Found. Phys.} \textbf{\bibinfo{volume}{18}},
  \bibinfo{pages}{95} (\bibinfo{year}{1988}).

\bibitem[{\citenamefont{Oldershaw}(1989{\natexlab{a}})}]{Oldershaw1989a}
\bibinfo{author}{\bibfnamefont{R.~L.} \bibnamefont{Oldershaw}},
  \bibinfo{journal}{International Journal of Theoretical Physics}
  \textbf{\bibinfo{volume}{28}}, \bibinfo{pages}{669}
  (\bibinfo{year}{1989}{\natexlab{a}}).

\bibitem[{\citenamefont{Oldershaw}(1989{\natexlab{b}})}]{Oldershaw1989b}
\bibinfo{author}{\bibfnamefont{R.~L.} \bibnamefont{Oldershaw}},
  \bibinfo{journal}{International Journal of Theoretical Physics}
  \textbf{\bibinfo{volume}{28}}, \bibinfo{pages}{1503}
  (\bibinfo{year}{1989}{\natexlab{b}}).

\bibitem[{\citenamefont{Dmitriyev}(1992)}]{Dmitriyev1992}
\bibinfo{author}{\bibfnamefont{V.~P.} \bibnamefont{Dmitriyev}},
  \bibinfo{journal}{Mechanics of Solids} \textbf{\bibinfo{volume}{26}},
  \bibinfo{pages}{60} (\bibinfo{year}{1992}).

\bibitem[{\citenamefont{Larson}(1998)}]{Larson1998}
\bibinfo{author}{\bibfnamefont{D.~J.} \bibnamefont{Larson}},
  \bibinfo{journal}{Physics Essays} \textbf{\bibinfo{volume}{11}},
  \bibinfo{pages}{524} (\bibinfo{year}{1998}).

\bibitem[{\citenamefont{Marmanis}(1998)}]{Marmanis1998}
\bibinfo{author}{\bibfnamefont{H.}~\bibnamefont{Marmanis}},
  \bibinfo{journal}{Phys. Fluids} \textbf{\bibinfo{volume}{10}},
  \bibinfo{pages}{1428} (\bibinfo{year}{1998}).

\bibitem[{\citenamefont{Dmitriyev}(1998)}]{Dmitriyev1998}
\bibinfo{author}{\bibfnamefont{V.~P.} \bibnamefont{Dmitriyev}},
  \bibinfo{journal}{Nuovo Cimento} \textbf{\bibinfo{volume}{111A}},
  \bibinfo{pages}{501} (\bibinfo{year}{1998}).

\bibitem[{\citenamefont{Zareski}(2001)}]{Zareski2001}
\bibinfo{author}{\bibfnamefont{D.}~\bibnamefont{Zareski}},
  \bibinfo{journal}{Found. Phys. Lett.} \textbf{\bibinfo{volume}{14}},
  \bibinfo{pages}{447} (\bibinfo{year}{2001}).

\bibitem[{\citenamefont{Dmitriyev}(2003)}]{Dmitriyev2003}
\bibinfo{author}{\bibfnamefont{V.~P.} \bibnamefont{Dmitriyev}},
  \bibinfo{journal}{Am. J. Phys.} \textbf{\bibinfo{volume}{71}},
  \bibinfo{pages}{952} (\bibinfo{year}{2003}).

\bibitem[{\citenamefont{Einstein}(1983)}]{Einstein1920}
\bibinfo{author}{\bibfnamefont{A.}~\bibnamefont{Einstein}},
  \emph{\bibinfo{title}{Aether and the Theory of Relativity, 1920, translated
  in 'Sidelights on Relativity', Dover}} (\bibinfo{year}{1983}).

\bibitem[{\citenamefont{Kostro}(2000)}]{Kostro2000}
\bibinfo{author}{\bibfnamefont{L.}~\bibnamefont{Kostro}},
  \emph{\bibinfo{title}{Einstein and the Ether}} (\bibinfo{publisher}{Apeiron
  4405, Rue St-Dominique Montreal, Quebec H2W 2B2 Canada,
  http://redshift.vif.com}, \bibinfo{year}{2000}).

\bibitem[{\citenamefont{Dmitriyev}(2004)}]{Dmitriyev2004}
\bibinfo{author}{\bibfnamefont{V.~P.} \bibnamefont{Dmitriyev}},
  \bibinfo{journal}{Meccanica} \textbf{\bibinfo{volume}{39}},
  \bibinfo{pages}{511} (\bibinfo{year}{2004}).

\bibitem[{\citenamefont{Tan}(2005)}]{Tan2005}
\bibinfo{author}{\bibfnamefont{Q.-M.} \bibnamefont{Tan}},
  \emph{\bibinfo{title}{Dimensional Analysis, in Chinese}}
  (\bibinfo{publisher}{University of Science and Technology of China Press,
  Beijing}, \bibinfo{year}{2005}).

\bibitem[{\citenamefont{Stokes}(1845)}]{Stokes1845}
\bibinfo{author}{\bibfnamefont{G.~G.} \bibnamefont{Stokes}},
  \bibinfo{journal}{Phil. Mag.} \textbf{\bibinfo{volume}{27}},
  \bibinfo{pages}{9} (\bibinfo{year}{1845}).

\bibitem[{\citenamefont{Stokes}(1849)}]{Stokes1849}
\bibinfo{author}{\bibfnamefont{G.~G.} \bibnamefont{Stokes}},
  \bibinfo{journal}{Trans. Camb. Phil. Soc.} \textbf{\bibinfo{volume}{9}},
  \bibinfo{pages}{1} (\bibinfo{year}{1849}).

\bibitem[{\citenamefont{Stokes}(1862)}]{Stokes1862}
\bibinfo{author}{\bibfnamefont{G.~G.} \bibnamefont{Stokes}},
  \bibinfo{journal}{Brit. Assoc. Reports} \textbf{\bibinfo{volume}{32}},
  \bibinfo{pages}{253} (\bibinfo{year}{1862}).

\bibitem[{\citenamefont{Eringen}(1975)}]{Eringen1975}
\bibinfo{author}{\bibfnamefont{A.~C.} \bibnamefont{Eringen}},
  \emph{\bibinfo{title}{Elastodynamics}} (\bibinfo{publisher}{Academic Press,
  New York}, \bibinfo{year}{1975}).

\bibitem[{\citenamefont{Landau and Lifshitz}(1986)}]{Landau-Lifshitz1986}
\bibinfo{author}{\bibfnamefont{L.~D.} \bibnamefont{Landau}} \bibnamefont{and}
  \bibinfo{author}{\bibnamefont{Lifshitz}}, \emph{\bibinfo{title}{Theory of
  elasticity, translated from the Russian by J.B. Sykes and W.H. Reid.}}
  (\bibinfo{publisher}{Pergamon, New York}, \bibinfo{year}{1986}).

\bibitem[{\citenamefont{Reiner}(1960)}]{Reiner1960}
\bibinfo{author}{\bibfnamefont{M.}~\bibnamefont{Reiner}},
  \emph{\bibinfo{title}{Lectures on Theoretical Rheology}}
  (\bibinfo{publisher}{North-Holland}, \bibinfo{year}{1960}).

\bibitem[{\citenamefont{Christensen}(1982)}]{Christensen1982}
\bibinfo{author}{\bibfnamefont{R.~M.} \bibnamefont{Christensen}},
  \emph{\bibinfo{title}{Theory of viscoelasticity : an introduction}}
  (\bibinfo{publisher}{Academic Press, New York}, \bibinfo{year}{1982}).

\bibitem[{\citenamefont{Joseph}(1990)}]{Joseph1990}
\bibinfo{author}{\bibfnamefont{D.~D.} \bibnamefont{Joseph}},
  \emph{\bibinfo{title}{Fluid Dynamics of Viscoelastic Liquids}}
  (\bibinfo{publisher}{Springer-Verlag, New York}, \bibinfo{year}{1990}).

\bibitem[{\citenamefont{Maxwell}(1868)}]{Maxwell1868}
\bibinfo{author}{\bibfnamefont{J.~C.} \bibnamefont{Maxwell}},
  \bibinfo{journal}{Phil. Mag.} \textbf{\bibinfo{volume}{35}},
  \bibinfo{pages}{129,185} (\bibinfo{year}{1868}).

\bibitem[{\citenamefont{Boltzamann}(1874)}]{Boltzamann1874}
\bibinfo{author}{\bibfnamefont{I.}~\bibnamefont{Boltzamann}},
  \bibinfo{journal}{Sitzungsber. Math. Naturwiss. Kl. Kaiserl. Akad. Wiss.}
  \textbf{\bibinfo{volume}{70}}, \bibinfo{pages}{275} (\bibinfo{year}{1874}).

\bibitem[{\citenamefont{Wang et~al.}(1982)\citenamefont{Wang, Xiong, and
  Huang}}]{Wang1982}
\bibinfo{author}{\bibfnamefont{R.}~\bibnamefont{Wang}},
  \bibinfo{author}{\bibfnamefont{Z.~H.} \bibnamefont{Xiong}}, \bibnamefont{and}
  \bibinfo{author}{\bibfnamefont{W.~B.} \bibnamefont{Huang}},
  \emph{\bibinfo{title}{Foundations of Theory of Plasticity, in Chinese}}
  (\bibinfo{publisher}{Science Press, Beijing}, \bibinfo{year}{1982}).

\bibitem[{\citenamefont{Oldroyd}(1950)}]{Oldroyd1950}
\bibinfo{author}{\bibfnamefont{J.~G.} \bibnamefont{Oldroyd}},
  \bibinfo{journal}{Proc. R. Soc., London, Ser. A}
  \textbf{\bibinfo{volume}{200}}, \bibinfo{pages}{523} (\bibinfo{year}{1950}).

\bibitem[{\citenamefont{Pao and Mow}(1973)}]{Pao-Mow1973}
\bibinfo{author}{\bibfnamefont{Y.-H.} \bibnamefont{Pao}} \bibnamefont{and}
  \bibinfo{author}{\bibfnamefont{C.-C.} \bibnamefont{Mow}},
  \emph{\bibinfo{title}{Diffraction of Elastic Waves and Dynamic Stress
  Concentrations}} (\bibinfo{publisher}{Crane, Russak \& Company Inc. in US},
  \bibinfo{year}{1973}).

\bibitem[{\citenamefont{Pao}(1983)}]{Pao1983}
\bibinfo{author}{\bibfnamefont{Y.-H.} \bibnamefont{Pao}}, \bibinfo{journal}{J.
  of Appl. Mech} \textbf{\bibinfo{volume}{12}} (\bibinfo{year}{1983}).

\bibitem[{\citenamefont{Long}(1967)}]{Long1967}
\bibinfo{author}{\bibfnamefont{C.~F.} \bibnamefont{Long}},
  \bibinfo{journal}{Acta Mech.} \textbf{\bibinfo{volume}{3}},
  \bibinfo{pages}{371} (\bibinfo{year}{1967}).

\bibitem[{\citenamefont{Lamb}(1932)}]{Lamb1932}
\bibinfo{author}{\bibfnamefont{H.}~\bibnamefont{Lamb}},
  \emph{\bibinfo{title}{Hydrodynamics}} (\bibinfo{publisher}{Cambridge
  University Press}, \bibinfo{year}{1932}), \bibinfo{edition}{6th} ed.

\bibitem[{\citenamefont{Kochin et~al.}(1964)\citenamefont{Kochin, Kibel, and
  Roze.}}]{Kochin1964}
\bibinfo{author}{\bibfnamefont{N.~E.} \bibnamefont{Kochin}},
  \bibinfo{author}{\bibfnamefont{I.~A.} \bibnamefont{Kibel}}, \bibnamefont{and}
  \bibinfo{author}{\bibfnamefont{N.~V.} \bibnamefont{Roze.}},
  \emph{\bibinfo{title}{Theoretical hydrodynamics, Translated from the fifth
  Russian ed.}} (\bibinfo{publisher}{Interscience Publishers, New York},
  \bibinfo{year}{1964}).

\bibitem[{\citenamefont{Yih}(1969)}]{Yih1969}
\bibinfo{author}{\bibfnamefont{C.-S.} \bibnamefont{Yih}},
  \emph{\bibinfo{title}{Fliud Mechanics}} (\bibinfo{publisher}{McGraw-Hill Book
  Company, New York}, \bibinfo{year}{1969}).

\bibitem[{\citenamefont{Wu}(1982)}]{Wu1982a}
\bibinfo{author}{\bibfnamefont{W.-Y.} \bibnamefont{Wu}},
  \emph{\bibinfo{title}{Fluid Mechanics, vol. 1, in Chinese}}
  (\bibinfo{publisher}{Beijing University Press, Beijing},
  \bibinfo{year}{1982}).

\bibitem[{\citenamefont{Faber}(1995)}]{Faber1995}
\bibinfo{author}{\bibfnamefont{T.~E.} \bibnamefont{Faber}},
  \emph{\bibinfo{title}{Fluid Dynamics for Physicists}}
  (\bibinfo{publisher}{Cambridge University Press, Cambridge},
  \bibinfo{year}{1995}).

\bibitem[{\citenamefont{Currie}(2003)}]{Currie2003}
\bibinfo{author}{\bibfnamefont{I.~G.} \bibnamefont{Currie}},
  \emph{\bibinfo{title}{Fundamental Mechanics of Fluids}}
  (\bibinfo{publisher}{Cambridge University Press}, \bibinfo{year}{2003}),
  \bibinfo{edition}{3rd} ed.

\end{thebibliography}


\end{document}